\documentclass[11pt]{article}
\usepackage{amsmath}
\usepackage{latexsym}
\usepackage{amssymb}

\newcommand{\ten}{\overline{\mathbf{10}}_{f}}
\newcommand{\eight}{\mathbf{8}_{f}}
\newcommand{\Th}{\Theta^{+}}

\let\cal\mathcal

\newcommand{\half}{\frac{1}{2}}

\newcommand{\beq}{\begin{equation}}
\newcommand{\eeq}{\end{equation}}
\newcommand{\ba}{\begin{array}}
\newcommand{\ea}{\end{array}}

\newcommand{\dd}{Q\!\!\!\!Q}
\newcommand{\ddd}{D\!\!\!\!D}

\title{A Perspective on Pentaquarks
\footnote{Plenary talk delivered by F. Wilczek for the European Physical Society,
Aachen, August 2003.}}
\author{Robert Jaffe and Frank
Wilczek\footnote{jaffe@mit.edu, wilczek@mit.edu}}
\begin{document}
\maketitle

\begin{abstract}
Recent discoveries of manifestly exotic, narrow baryon resonances present a fundamental challenge
for our theoretical understanding of low-energy QCD.  This is a brief survey of their interpretation,
emphasizing the possibility that diquark correlations are centrally involved.  Many predictions and
suggestions for future directions follow from that idea.
\end{abstract}

Recently several experimental groups have reported observation of a
new, manifestly exotic (B=1, S=1) baryon resonance $\Th$(1540)
\cite{Nakano:2003qx, Barmin:2003vv,
Stepanyan:2003qr, Lorenzon, Kubantsev, Kubarovsky:2003fi}.  The
resonance seems unusually narrow ($\Gamma<$~15 Mev) for a particle in
this mass range with open channels to strong decay.  Shortly after the
conference the NA49 group at CERN announced evidence for an additional
narrow ``cascade'' exotic (B=1, Q=S=-2) $\Xi^{--}$ and also $\Xi^-,
\Xi^0$ with masses close to 1860 Mev \cite{Alt:2003vb, jlab}. 
(Particles of this kind were predicted in my talk, at the time it was
delivered.)  Since the simplest quark assignments consistent with the
quantum numbers of $\Th$ and $\Xi^{--}$ are ($udud\bar{s}$) and
($dsds\bar{u}$) respectively, these particles have come to be called
``pentaquarks''.

The discoveries of manifestly exotic particles, which have been sought for
decades, clearly open a new chapter in strong interaction physics.  How
fundamental are they?  What can we hope to learn?

Although the discoveries are striking, I don't think they are so
peculiar as to require introducing
new interactions or modifying QCD as the basic theory of the strong
interaction.  We don't know how to calculate the consequences of QCD
accurately at low energy, in general.  But we know that numerical
solution of the equations works brilliantly in the dozen or so cases
where it has been done; we know that the perturbation theory, with
asymptotic freedom, works brilliantly at high energy; and we know that
the basic theory is so tight conceptually that we don't have sensible
ways to modify it, even if we wanted to.  So it seems unlikely that
we'll be driven to modify the basic equations of QCD or the standard
model, and no one has suggested this seriously.

On the other hand these discoveries do offer us a golden opportunity to
sharpen and expand our understanding of QCD itself.  In that context, I
will argue, they could well have major impact upon several of the most
fundamental unresolved issues.

Let me remind you how opportunistic, or I might say schizoid, our
conventional, pragmatic approach to hadron dynamics is.

When we do precision work, we use honest quantum field theory: either
straight numerical integration of the full equations using computers
and the techniques of lattice gauge theory, or perturbation theory
when justified by asymptotic freedom.  In relativistic quantum
field theory the states of definite energy, such as protons and pions
in QCD, appear to be very complicated.  If we insist on trying to
constructing them as states in Fock space (which, ultimately, we can't
really do) we would expect to find coherent superpositions of states
containing various numbers of quark-antiquark pairs and gluons.  The
proton structure function in deep inelastic scattering, if interpreted
in constituent language, bears this out: there are in fact an infinite
number of gluons and quark-antiquark pairs lurking at small $x$.  So,
in a different way, does treatment of pions as Nambu-Goldstone bosons
of chiral symmetry breaking: these are collective modes, constructed
as waves within a condensate of tachyonic quark-antiquark pairs.

Yet when we turn to practical spectroscopy, we find the literature is
dominated by simple models based on essentially non-interacting quarks
living in a mean field or bag.  Singlet quark-antiquark pairs, gluons,
and correlations are censored out or integrated into effective
constitutent degrees of freedom.  This sort of na\"{\i}ve quark model is
easy to use and it organizes a lot of data pretty successfully, which
is why it's useful and popular.  But it's a dead end.  Consideration
of pentaquarks brings some serious shortcomings of the na\"{\i}ve quark
model into sharp focus.  By confronting them we may well reach a new
level of understanding, and elucidate some old and important problems
in strong interaction physics that so far have been ``bypassed'', rather
than addressed, in QCD.


\section {{\large Basic Modeling Choices}}

I'm going to be advocating a particular interpretation of the emerging physics, and
I admit up front that it is by no means established or universally
accepted.  It does have the virtue of being readily falsifiable, so I
expect that this talk will come to look either prescient or hopelessly
misguided within a few years.  In fairness I should say a few words about
the leading alternatives, and how they can be distinguished.

One's first instinct might be to model the $\Th$ as weakly bound
kaon-nucleon molecule.  QCD allows $KN$ to couple to $q^{4}\bar
q$ configurations with wave functions that differ markedly in space,
color, and spin from the $KN$ scattering state.  If we want to avoid
introducing such effects explicitly, we have no choice but to describe
$KN$ scattering in the region of the $\Th$ in terms of
non-relativistic potential scattering.  Let's see where that leads.
Attractive forces in the $s$-wave do not generate resonances.  If they
are attractive enough they produce bound states, if not, they modulate
the phase of the scattering wave function smoothly.  In higher partial
waves attractive forces can generate resonances through interplay
between attraction and the repulsive angular momentum barrier.  The
mass and width of the resonance are related through the range and
depth of the potential.  For a simple attractive potential with range
1 fermi, the width of a $P$-wave resonance 100 MeV above threshold is
more than 175 MeV. To obtain a width of order 10 MeV we would have to
adjust the range to about $0.05$ fermi!  Thus the observed parameters
of the $\Th$ can't be fit without bringing in strong, short-range
attraction, which would be inconsistent with everything we know about
QCD.

The chiral soliton or Skyrme model has played an important role in
stimulating the search for exotics \cite{csm}.  Indeed, the initial experiments
leading to the discovery of $\Th$ were stimulated by the suggestions
of Diakonov, Petrov, and Polyakov.  Their procedures have been
criticized on theoretical grounds, however, and their original
prediction for the mass of $\Xi^{--}$ was far too large.  There is
considerable flexibility in the model, and it might be possible to
accommodate the existing pentaquark phenomenology.  But whatever
it may say about pentaquarks the
model touches only a very small subset of the resonances you find in
the Particle Physics Booklet (it does not describe the $\rho$, for example), and seems
to rely heavily on inessential aspects of QCD (that is, without very
light quarks the Skyrme model ceases to exist, but most of QCD hardly
changes).  My feeling is that an approach based more directly on
fundamental degrees of freedom in QCD is more likely to be fruitful
--- but who knows?

Now let's consider what the na\"{\i}ve (uncorrelated) quark model says
about pentaquarks.  In the uncorrelated quark model, in which all the
quarks are in the ground state of a mean field, the ground state of
$q^{4}\bar q$ has \emph{negative} parity.  The full $SU(3)_{f}\times
SU(2)_{\rm spin}$ content of this multiplet and bag model estimates of
masses and decay couplings can be found in a 1979 paper by
Strottman \cite{Strottman:qu}.
This picture in no way explains the narrowness of the observed
pentaquarks.  There are very many states in flavor $\mathbf{1}_{f}$,
$\eight$, $\ten$, $\mathbf{27}_{f}$, {\it etc.\/}.  The lightest have
quark content $uudd\bar d$ and $uudd\bar u$ and would be expected to
lie below the $\Th$.  There is no evidence for a $\half^{-}$ nucleon
in this well explored region of the non-strange baryon spectrum.
Furthermore, all the known light negative parity baryons are well
described as orbital excitations of $q^{3}$.  Altogether, the
uncorrelated model of pentaquarks appears to be extremely problematic,
on empirical grounds.

It is worth re-emphasizing that the uncorrelated quark model, where all the
quarks are in the ground state of a mean field, assigns the ground state of
$q^{4}\bar q$ (and therefore $\Th$ and $\Xi^{--}$), \emph{negative} parity. Both
the correlated quark model I'll be advocating and the chiral soliton model
predict positive parity.  The experimental verdict on this crucial issue is
not yet in.


\section{\large{Diquarks}}

Attraction between quarks in the color $\mathbf{\overline{3}}_{c}$ channel
has profound roots in microscopic QCD. Indeed, by bringing quarks together
in this channel one halves the magnitude of their effective charge, and
thus largely cancels the associated field energy.  More particularly, the
strongest attraction is in the flavor antisymmetric $J^P = 0^+$ channel.
This channel is also favored by instanton-mediated interactions.  At high
baryon number density one can calculate rigorously that this is the channel
in which quark Cooper pairs condense, to produce color superconductivity
and color-flavor locking.

From now on, when I refer to diquarks I shall always have this color and
flavor $ {\overline{3}_{f}}$, $J^P = 0^+$ channel in mind and use
the symbol $\dd$ for it.

In vacuum of course the quark-antiquark color singlet flavor singlet
channel is even more attractive than the diquark.  Condensation in the $\bar{q}q$ channel, which
drives chiral symmetry breaking, supersedes condensation in the diquark
channel.  Nevertheless some suggestive signs of diquark attraction have occasionally
been discerned.  Perhaps the most compelling, and in any case the most
immediately relevant concern exotic spectroscopy:

\begin{itemize}

\item The nearly ideally mixed nonet of $J^{\Pi}=0^{+}$ meson
including $f_{0}(600)$, $\kappa(800)$, $f_{0}(980)$, and $a_{0} (980)$
have always posed classification problems for conventional quark
models.  In particular, their mass spectrum is inverted, and there is
an entire additional nonet of scalar mesons in the 1100--1500 MeV
range, where $q\overline q$ mesons would be expected to lie.  It is
tempting to classify the light nonet as $qq\bar q \bar q$, with the
quarks and antiquarks organized into $\dd$ and $\overline{\dd}$
respectively.

\item The observed {\it absence\/} of manifestly exotic $qq\bar q\bar q$
mesons is a remarkable fact in itself.  It is explained by the correlation
of quarks into diquarks, since the product of diquark and antidiquark
produces the same flavor quantum numbers as quark times antiquark.  Without
correlations, of course, many exotic representations are possible.

\item There's a similar story for light-quark baryons made from four quarks
and an antiquark.  {\it A priori\/} there are all sorts of possibilities
for light exotics.  But the only exotics observed belong (presumably) in
the $\Th$ antidecuplet.  Where are the others?  Without correlations, it's
a puzzle.  Diquark correlations single out the observed antidecuplet uniquely.
Since diquarks are $SU(3)$-flavor antitriplets, the only way to make an
exotic out of two diquarks and an antiquark is to combine the diquarks
symmetrically in flavor, [$\overline{3}_{f} \otimes \overline{3}_{f}]_{\cal
S} = \overline{6}_{f}$, and then couple the antiquark.  The flavor content
of the resulting $q^{4}\overline q$ states is then $\bar 6_{f}\otimes \bar
3_{f}= 8_{f}\oplus \overline{10}_{f}$.  Note that the
antidecuplet comes together with an octet, which should mix when possible
to produce pure strange quark content (ideal mixing).

\item Other approaches to exotic spectroscopy predict a much richer
spectrum of exotics including $27_{f}$ and $35_{f}$ multiplets.  A
notable difference is the absence in the diquark picture of an
isovector analog of the $\Theta^{+}(1540)$, with $S=+1$ and charges
$Q=0$, $1$, and $2$.  This state occurs in the $27_{f}$ and other
exotic multiplets, but not in the $\overline{10}_{f}$.  Its
occurrence, with low mass, appears to be a robust prediction of chiral
soliton models \cite{mp}.  Targeted searches have come up
empty \cite{empty}.
	
\end{itemize}


\section{\large{Pentaquarks}}

My original discussion of detailed pentaquark phenomenology, which was
essentially a sketch of \cite{Jaffe:2003sg}, is already out of date due to the discoveries reported by
NA49 --- which are broadly consistent with it, but of course more
specific and richer in detail.  Since the situation is developing
rapidly, here I will only mention a few salient points.  For more on
the interpretation of NA49, including suggestions for additional
observables, see \cite{Jaffe}.

The exotic antidecuplet baryons should have spin-parity $1/2^{+}$ and
be accompanied by nearby states with
$J^{\Pi}=3/2^{+}$ \cite{Jaffe:2003sg,Dudek:2003xd}:
$[\dd\otimes \dd]_{\cal S}$ must be in the $p$-wave to satisfy Bose
statistics.  This $\ell=1$ system can couple to the antiquark to give
either $J^{\Pi}= {3/2}^{+}$ or $ {1/2}^{+}$.  At present both
possibilities are open for the cascades found by NA49.

The mass splittings of the $[\dd\otimes\dd]_{\cal S}\otimes \bar q $
octet and antidecuplet baryons, computed to first order in $m_{s}$,
yield a spectrum discussed in detail in 
\cite{Jaffe:2003sg}.  We would like to identify the
$\Th$ with the $\dd\dd\bar q$ state $  [ud]^{2}\bar s $.
The narrowness of the physical $\Th$ can be explained by the
relatively weak coupling of the $K^{+}n$ continuum to the
$[ud]^{2}\bar s$ state from which it differs in color, spin and
spatial wave functions.

$N([ud]^{2}\bar d)$ is the
lightest particle in the $\eight+\ten$.  It has the quantum number of the
nucleon.  It is tempting to identify this state with the otherwise
perplexing Roper resonance, the $N(1440)\ P_{11}$, which has defied
classification for decades.  The $N(1440)$ is much broader
than the $\Th$.  Of course the internal structure and group-theoretic
properties of $\Th$ and $N(1440)$ are quite different, and the $N(1440)$
can mix with the ordinary nucleon.

Most remarkably, we expect \emph{two} multiplets of cascades for each
spin.  These are an $I=3/2$ quartet arising from the decuplet, which
includes the manifestly exotic $\Xi^{+}(uuss\bar d)$ and $\Xi^{--}
(ddss\bar u)$; and an $I=1/2$ doublet with charges 0, -1.  This is
important because, as argued in 
\cite{Jaffe}, it is difficult to accommodate the NA49 observations
using an antidecuplet alone.  Because these states differ in isospin,
mixing between them should be quite small, barring extreme accidental
degeneracy.

Charm and bottom analogues of the $\Theta([ud][ud]\bar s)$ with quark
content $[ud][ud]\bar c$ and $[ud][ud]\bar b$ might be stable against
strong decay.  The strong decay thresholds for these states depend on
the corresponding pseudoscalar meson masses, which grow like the
square root of the quark masses.  Thus, for example, the threshold for
$\Theta^{0}_{c} ([ud][ud]\bar c) \to pD^{-}$ is relatively higher than
the threshold for $\Theta^{+}_{s}(uudd\bar s)\to nK^{+}$ 
\cite{Jaffe:2003sg}.

\section{\large{{Future Directions}}}

Clearly, the first order of business must be to clarify and solidify
the experimental situation.  There is a host of predictions to check:
positive parity; a crowded 4-component spectrum of light pentaquarks
including nearby spin 1/2 and 3/2 flavor octet and antidecuplet
multiplets; and narrow, possibly strongly stable exotics containing
heavy antiquarks.  Although I call this spectrum
``crowded'' you should recognize that it is far sparser in
exotics than what is suggested by the uncorrelated quark model, or in
implementations of the chiral soliton idea.

On the theoretical side, one important direction is to bring the power
of lattice gauge theory to bear on these issues.  Here the most
obvious challenge is to find the pentaquarks.\cite{lattice} If the
diquark picture is on the right track, that won't be entirely trivial
to do, for two reasons.  First, it is complicated to construct sources
that are well matched to pentaquarks.  In particular, they must have a
very particular color structure, and rather complicated spatial
structure reflecting the relative p-wave.

In the diquark of interest the quarks are coupled antisymmetrically in
color, spin, and flavor, to the $\overline 3_{f}$, $\overline
3_{c}$, $J=0$ representations.  Let $q^{ai}$ be a quark Dirac
field.  $a$ is an $SU(3)_{f}$ index; $i$ is an $SU(3)_{c}$ index.
Then define,
\begin{equation}
	\dd_{ck}= \epsilon_{abc}\epsilon_{ijk}q^{ai}i\sigma_{2}q^{bj}
\label{eq1}
\end{equation}
where $\sigma_{2}$ is the usual Dirac matrix.  Then $\dd_{ck}$ is
in the representations required.

In our pentaquark model two diquarks are coupled antisymmetrically in color, to a
$3_{c}$, and symmetrically in flavor, to a $\overline 6_{f}$.  This
double-diquark carries a covariant $3_{c}$ label, $k$, and a pair
of covariant $\overline 3_{f}$ labels, $\{ab\}$.  Obviously
\begin{equation}
	{\cal S}_{\{ab\}\{cd\}}=\frac{1}{2}(\delta_{ac}\delta_{bd}+
	\delta_{ad}\delta_{bc})
	\label{eq2}
\end{equation}
couples two antiquarks with labels $c$ and $d$ to the symmetric
representation labeled by the symmetric pair $\{ab\}$.  So the pair of
diquarks, properly coupled, is
\begin{equation}
	\ddd^{k}_{\{ab\}}= \epsilon^{ijk}{\cal S}_{\{ab\}\{cd\}}
	\dd_{ci}\dd_{dj}
	\label{eq3}
\end{equation}
As it stands, this operator is identically zero when all the quarks are in
the same eigenmode of some mean field (i.e. if the Dirac field
is replaced by a single creation operator), as a consequence of Fermi
statistics.  Therefore we must introduce a derivative, effectively giving
the diquarks one unit of relative angular momentum
\begin{equation}
	\ddd^{k,\mu}_{\{ab\}}= \epsilon^{ijk}{\cal S}_{\{ab\}\{cd\}}
	\left(\dd_{ci}(D^{\mu}\dd_{dj})-(D^{\mu}\dd_{ci})\dd_{dj}
	\right)
	\label{eq3}
\end{equation}
Note the minus sign between the two terms, chosen to produce a unit of
relative angular momentum between the diquark pairs.  The covariant
derivative, $D^{\mu}$, is in the $\overline 3$ representation of color
in order that the field $(D^{\mu}\dd)$ transforms as an $\overline 3$
\begin{equation}
	D^{\mu}=\partial^{\mu}-ig\lambda^{\dagger}_{\ell}A^{\mu \ell}
\end{equation}

The coupled diquarks transform like a Lorentz vector.  This vector can
be coupled to an antiquark to form baryon fields with angular momentum
$1/2$ or $3/2$.  It is simple to construct the appropriate Lorentz
representations out of general vector, $V^{\mu}$, and a Dirac spinor
$q$.  The spin $1/2$ field is $V^{\mu}\gamma_{\mu}q$ and the spin
$3/2$ field is $V^{\nu}\sigma_{\mu\nu}q$.  The only complication for
us is that these baryons are built from an \emph{antiquark} and two
diquarks, so the correct forms are $B=V^{\mu}\gamma_{\mu}q_{\cal C}$
and $B_{\mu}=\sigma_{\mu\nu}V^{\nu}q_{\cal C}$, where $q_{\cal
C}={\cal C}\bar q^{\rm T}=i\gamma_{2}q^{*}$.  Altogether,\cite{thanks}
\begin{eqnarray}
	B_{\{ab\}d} =&\ddd^{k,\mu}_{\{ab\}}\gamma_{\mu} (q_{\cal C})_{dk}
	 &=
	 \epsilon^{ijk}{\cal S}_{\{ab\}\{ef\}}
	\left(\dd_{ei}(D^{\mu}\dd_{fj})-(D^{\mu}\dd_{ei})\dd_{fj}
	\right)\gamma_{\mu} \gamma_{5}\,(q_{\cal C})_{dk} \nonumber\\
	B_{\{ab\}d,\mu} =&\ddd^{k,\nu}_{\{ab\}}\sigma_{\mu\nu} (q_{\cal C})_{dk}
	 &=
	 \epsilon^{ijk}{\cal S}_{\{ab\}\{ef\}}
	\left(\dd_{ei}(D^{\nu}\dd_{fj})-(D^{\nu}\dd_{ei})\dd_{fj}
	\right) \sigma_{\mu\nu}(q_{\cal C})_{ck}
\end{eqnarray}
Finally, the antidecuplet is projected out of these fields by
symmetrizing over all the $SU(3)_{f}$ labels, $a$, $b$, and $c$.
These are the sources we expect to couple well to pentaquarks.
They bear little resemblence to the sources used in the first
attempts to examine pentaquark spectra on the lattice.

Second, the diquark attraction is most effective for very light
quarks, and these are difficult to handle numerically.  Even the usual
pseudoscalar mesons remain a challenge --- one might say an
embarrassment --- for lattice gauge theory, for the same reason.  One
can simplify life somewhat, and also address an intrinsically
interesting case, by specializing the antiquark to be a fixed color
source.

These are significant technical problems, but I'm sure
that ingenious people using powerful computers will overcome them.

A more open-ended challenge for lattice gauge theory is to look for diquark
correlations more broadly, in the various contexts they have been
suggested.  By probing how strongly the light scalars, or for that matter
nucleons, couple to different kinds of sources we can see if there is
evidence for significant diquark content, for example.

On the more phenomenological side, it should be useful and instructive to
consider expanding nonrelativistic quark or (better) bag models to include
fundamental diquark degrees of freedom, with appropriate interactions, to
see whether several different sectors (e.g., the scalar nonet and its heavy
cousins light and heavy pentaquarks) can be described in a common
semi-quantitative framework.

It is also tempting to speculate more freely.  Is the reason for ``hard
core'' repulsion between nucleons the short-range repulsion between the
diquarks they contain?  Is the reason for the $\Delta I = 1/2$ rule
that the piece of the nonleptonic weak Hamiltonian that contains
diquark quantum numbers is enhanced?  Can we put the study of quark
correlations using electroweak scattering probes, which already has
been interpreted as providing evidence for diquarks in nucleons, on a
rigorous footing?

Finally, I'd like to make two brief remarks of a theoretical nature:

\begin{enumerate}

\item By gauging flavor SU(2) strongly, one could enforce strong diquark
correlations.  So there is a logically consistent limit in which the
diquark picture is manifestly appropriate.  The remaining question, of
course, how much of its characteristic structure survives as we go away
from that limit, by decreasing the extra gauge coupling.  That might be an
interesting avenue to explore numerically.

\item Important diquark correlations would seem to be against the spirit of
conventional large N approaches to QCD. The crucial circumstance that two
quarks can  lower their color charge drastically by coming together is
special to N=3.

\end{enumerate}

We can look forward to a vigorous dialogue among traditional laboratory
experiments, numerical quasi-experiments, and theoretical explorations.  It
should be great fun, and I'll be surprised if the outcome is not a better
understanding of how the murkiest part of our fundamental theory of matter
- i.e., low-energy QCD - really works.


\end{document}